\documentclass[pra,twocolumn,showpacs,preprintnumbers,amsmath,amssymb,tightenlines,epsfig]{revtex4}

\usepackage{graphicx}% Include figure files
\usepackage{dcolumn}% Align table columns on decimal point
\usepackage{bm}% bold math

\begin{document}
\preprint{Version: submitted \today}
\title{Control of an atom laser using feedback}
\author{S. A. Haine, A. J. Ferris, J. D. Close and J. J. Hope}
\affiliation{Australian Centre for Quantum Atom Optics, Department of Physics, 
Australian National University, ACT 0200, Australia}

\email{Simon.Haine@anu.edu.au}

\begin{abstract}
A generalised method of using feedback to control Bose-Einstein condensates is introduced.  The condensates are modelled by the Gross-Pitaevskii equation, so only semiclassical fluctations can be suppressed, and back-action from the measurement is ignored.  We show that for any available control, a feedback scheme can be found to reduce the energy while the appropriate moment is still dynamic.  We demonstrate these schemes by considering a condensate trapped in a harmonic potential that can be modulated in strength and position.  The formalism of our feedback scheme also allows the inclusion of certain types of non-linear controls.  If the non-linear interaction between the atoms can be controlled via a Feshbach resonance, we show that the feedback process can operate with a much higher efficiency.
\end{abstract}


\maketitle

\section{Introduction}
In recent years, we have seen the first examples of the atom laser, a device similar to the optical laser, providing a coherent, Bose-condensed output beam \cite{Ketterle,Kasevich,Phillips,Esslinger,Martin}. The development of the atom laser past the demonstration stage, particularly the development of the pumped atom laser, is an important goal in atom optics. In many applications, it is the high spectral flux and coherence provided by a pumped laser that is critical. Pumping, however, is difficult to implement with atoms and can lead to classical noise that far exceeds the suppression of quantum noise, or line narrowing, that would be expected from a pumped system. This paper presents a method of suppressing the classical noise on a pumped atom laser beam by feedback to the condensate, with the aim of achieving quantum noise limited operation.

As with light, the matter waves from an atom laser can be coherently reflected, focused, beam split, and polarised \cite{Esslinger2}.  These are the basic operations performed in all optics experiments and through these operations every linear, non-linear and quantum optics experiment has its analogue when performed with atoms. Although bosonic atoms and photons both exhibit Bose-stimulated scattering that is fundamental to laser operation \cite{Ketterle4,Ketterle5}, there are significant and interesting differences. The free space dispersion relation for atoms leads to spatial broadening of pulses in vacuum. Atoms interact with each other and display non-linear effects in the absence of another medium. Atoms display far more complex polarisation states, move slowly and can be readily produced with wavelengths much shorter than is available from an optical laser. These are ideal properties in many precision measurement and quantum information applications. 

The present state of the art in atom lasers is an unpumped Bose-Einstein condensate (BEC) that serves as a source for a propagating matter wave beam. Atoms are outcoupled from the condensate via an RF, or a Raman transition that coherently flips a trapped spin state to an untrapped state. There have been several experimental investigations of the properties of atom laser beams. Both temporal and spatial coherence have been measured, and it has been demonstrated that RF outcoupling preserves the coherence of the condensate \cite{Koehl,Bloch}. The beam divergence has been measured \cite{Aspect}, and there has been one real time measurement of the flux of an atom laser beam \cite{Esslinger3}. The four wave mixing experiments performed by the NIST group were the first experiments to exploit the inherent nonlinearity of atoms in a controlled fashion and, furthermore, demonstrated that the Raman outcoupling process also preserved the coherence displayed by the condensate \cite{Phillips2}. There have been two early experiments reporting squeezing in atom laser beams \cite{Kasevich2,Ketterle2}. Despite these pioneering experiments, there is a significant amount of development needed if the atom laser is to become a generally applicable and useful tool in quantum atom optics.

High spectral flux in optical lasers is generated through a competition between a depletable pumping mechanism that operates at the same time as the damping. The linewidth of a pumped laser is much narrower than the linewidth  of the cavity determined by the cavity lifetime. In a pumped laser, there is Bose enhancement of the scattering rate into the lasing mode resulting in line narrowing \cite{Wiseman, Holland,Hope00,Ashton2003}. The line narrowing, or suppression of quantum noise associated with pumping an atom laser, is a very desirable but as yet unrealised property.  Quantum field theory is required to calculate the quantum noise limited linewidth of an atom laser with interactions. Wiseman and Thomsen have studied the quantum noise on an atom laser beam outcoupled from a single mode condensate and have included feedback in their model. Atomic collisions turn number fluctuations into phase fluctuations significantly increasing the linewidth. A continuous QND feedback scheme can be used to cancel this linewidth broadening \cite{WiseThom}. 
It would be difficult to treat both quantum and classical noise in the same model, as the full quantum field theory is only tractable in the limit of a few modes, whereas the classical noise is intrinsically a multimode effect. 
There is no guarantee that a real atom laser would operate at the quantum noise limit, and it is likely that we must design pumping schemes very carefully and use feedback to approach the quantum noise limit. It is this goal that motivates the present work. 

The classical noise on a pumped atom laser can be studied with multimode semiclassical Gross-Pitaevski (GP)  models \cite{Nick}.  In a recent paper, it was shown that an atom laser pumped by a non mode-selective pumping scheme was unstable below a critical value of the scattering length leading to significant classical noise on the outcoupled beam \cite{Simon1}. It would seem sensible to adjust the scattering length via a Feschbach resonance to a suitably large value to stabilise the atom laser and reduce classical noise. Quantum and classical noise scale oppositely with scattering length, quantum noise increasing with scattering length and classical noise decreasing. The solution is to operate at low scattering length and either use mode selective pumping to stabilise the laser, or to suppress classical noise by feeding back to the condensate. Mode selective pumping would appear to be difficult to implement, and it is the second option, suppression of classical noise by feedback, that we investigate here.  

Any realistic feedback scheme will require a detector to measure classical noise, and a control to feedback to in order to suppress the motion of the condensate. The entire feedback loop must have sufficient bandwidth and must be minimally destructive. The design of minimally destructive detectors for real time measurement and feedback to stabilise an atom laser was discussed in two recent papers \cite{Jessica,JoeJohn}. In the present work, we have chosen to feedback to realistic controls provided by the magnetic trap to ensure straightforward implementation in an experiment. 

\section{Control of a condensate} 
The choice of an effective feedback scheme is largely determined by the available methods of controlling that system.  For a BEC, these controls can correspond either to perturbations in the trap potential, or changes in the interactions between the atoms.  We examine the feedback scheme required to control a BEC in three dimensions in an arbitary potential.  We model the system by the Gross-Pitaevskii equation.  We assume that it is possible to control a set of external potentials $\sum_{i}a_{i}(t)f_{i}(\bf{r})$ and spatially dependent nonlinear interaction strengths $\sum_{j}b_{j}(t)g_{j}(\bf{r})$ with time dependent amplitudes. With the feedback switched on, the equation of motion is:
\begin{equation}
i\hbar\frac{d\psi(\mathbf{r}, t)}{dt} = \hat{H}\psi(\mathbf{r}, t) \label{eom}
\end{equation}
with
\begin{eqnarray}
\hat{H} &=& \hat{H}_0 + \sum_{i}a_{i}(t)f_{i}(\mathbf{r})+ \sum_{j}b_{j}(t)g_{j}(\mathbf{r})|\psi|^2 \\
\hat{H}_0 &=& \hat{T} +V_0(\mathbf{r}) +U_0|\psi|^2, \quad \hat{T} = \frac{-\hbar^2}{2m}\nabla^2
\end{eqnarray}
The $a_i(t)$'s and $b_i(t)$'s are the set of controls used to manipulate the potentials.  We consider a condensate initially evolving under the Hamiltonian $\hat{H}_0$.  Unless the system is initially in the ground state, we want to reduce the energy, given by: 
\begin{equation}
E_{0}(\psi) = \langle \hat{T} +V_0 \rangle +\frac{1}{2}\langle U_0|\psi|^2 \rangle \label{Energy0}
\end{equation}
Where the angle brackets denote $\langle \hat{q} \rangle = \int \psi^{*} \hat{q}\psi d^3\bf{r}$, and the integral is over all space. By switching the feedback on, and then switching it back off again at some later time, we will typically have altered the value of $E_0$. It is important to note that in the presence of feedback, $E_0$ does not represent the instantaneous energy, but the energy that the system would have if the feedback were to be suddenly switched off at that time. The rate of change of $E_0$ while the feedback is switched on is:
\begin{equation}
\dot{E_0} = \int \dot{\psi}^{*}(\hat{T} +V_0)\psi +\psi^{*}(\hat{T} +V_0)\dot{\psi} +\frac{U_0}{2}\frac{d}{dt}|\psi|^4 d^3\mathbf{r}\label{E0dot}
\end{equation}
Where the dot $\dot{}$ denotes differentiation with respect to $t$. Using equation (\ref{eom}) in equation (\ref{E0dot}) and the fact that $\hat{H}$ is Hermitian gives: 
\begin{eqnarray}
\dot{E_0} &=& \frac{i}{\hbar}\langle \bigl[\hat{H}, \quad \hat{T} +V_0\bigr] \rangle + \frac{U_0}{2}\frac{d}{dt} \int |\psi|^4d^3\mathbf{r} \nonumber
\end{eqnarray} 
Using the divergence theorem gives
\begin{eqnarray}
&=& \frac{-i\hbar}{2m}\int \sum_{i}a_{i}f_{i}(\mathbf{r})\Bigl(\psi^{*}\nabla^2\psi -\psi\nabla^2\psi^{*} \Bigr) d^3\mathbf{r} \nonumber \\
&-&\frac{i\hbar}{2m}\int \sum_{j}b_{j}g_{j}(\mathbf{r})|\psi|^{2}\Bigl(\psi^{*}\nabla^2\psi -\psi\nabla^2\psi^{*}\Bigr)d^3\mathbf{r} \nonumber \\
&=& - \sum_{i}a_{i}(t)\dot{\bigl< f_{i}(\mathbf{r}) \bigr>} - \frac{1}{2}\sum_{j}b_{j}(t)\dot{\bigl< g_{j}(\mathbf{r})|\psi|^2 \bigr>}
\end{eqnarray}
It can be seen that setting $a_i(t) = c_i\dot{\langle f_i(\mathbf{r})\rangle}$ and $b_i(t) = u_i\dot{\langle g_i(\mathbf{r})|\psi|^2 \rangle}$, where $c_i$ and $u_j$ are positive constants, so that 
\begin{equation}
\dot{E_0} =  - \sum_{i}c_{i}\dot{\Bigl< f_{i}(\mathbf{r}) \Bigr>}^2 - \frac{1}{2}\sum_{j}u_{j}\dot{\Bigl< g_{j}(\mathbf{r})|\psi|^2 \Bigr>}^2,
\end{equation}
will always reduce $E_0$ while there are oscillations present in the appropriate moments of the condensate.
This is an important result as it illustrates a general scheme to reduce the energy from the condensate depending on the available controls. In the following sections, we demonstrate applying this feedback scheme to particular examples. In section \ref{section3}, we investigate how we can use feedback to control a linear ($U_0 = 0$) system in a harmonic potential. In section \ref{section4}, we demonstrate control of a Bose-Einstein condensate in a harmonic potential.

\section{Using the feedback scheme to control a linear harmonic oscillator} \label{section3}
\subsection{Harmonic oscillator with linear controls}
We now consider the specific example of the linear ($U_0=0$) Schr\"{o}dinger equation in one dimension with a harmonic potential, ie: $V_0 = \frac{1}{2}x^2$ (in harmonic oscillator units where $x$ is measured in units of the length $\sqrt{\frac{\hbar}{m\omega}}$, time $t$ is measured in units of the time $\omega^{-1}$ and energy is measured in units of $\hbar \omega$, where $\omega$ is the harmonic trapping frequency). We use as our controls the position of the minimum of the potential, and the strength of the potential.  This system is a good model of a BEC in either a magnetic or an optical trap, which are both approximately harmonic near the potential minimum, and can be modulated in intensity. The equation of motion is then
\begin{equation}
i\dot{\psi} = (\hat{T} + V_0 + a_1(t)x +a_2(t)x^2)\psi \label{HOeom}
\end{equation}
Setting $a_1(t) = c_1\dot{\langle x \rangle}$ and $a_2(t) = c_2\dot{\langle x^2 \rangle}$ in accordance with our feedback scheme gives
\begin{equation}
\dot{E_0} =  - c_{1}\dot{\langle x \rangle}^2 - c_{2}\dot{\langle x^2 \rangle}^2 
\end{equation}
This will guarantee that $E_0$ is always reduced while there are fluctuations in $\langle x \rangle$ and $\langle x^2 \rangle$, but the rate can be optimized  by carefully selecting the value $c_1$ and $c_2$.  We can calculate a dynamical equation for $\langle x \rangle$ using Ehrenfest's theorem \cite{griffiths}
\begin{equation}
\ddot{\langle x\rangle} = -\Bigl< \frac{\partial V(x, t)}{\partial x}\Bigr> = -(1 + 2a_2(t))\langle x \rangle - c_1\dot{\langle x \rangle}
\end{equation}
This is mathematically identical to a classical damped harmonic oscillator. Critical damping will occur when $c_1 = 2\sqrt{1+2a_2}$. The dynamic equation for $\langle x^2 \rangle$ isn't a simple linear harmonic oscillator, so we found an appropriate magnitude of $c_2$ numerically.  

\begin{figure}
\includegraphics[width=\columnwidth]{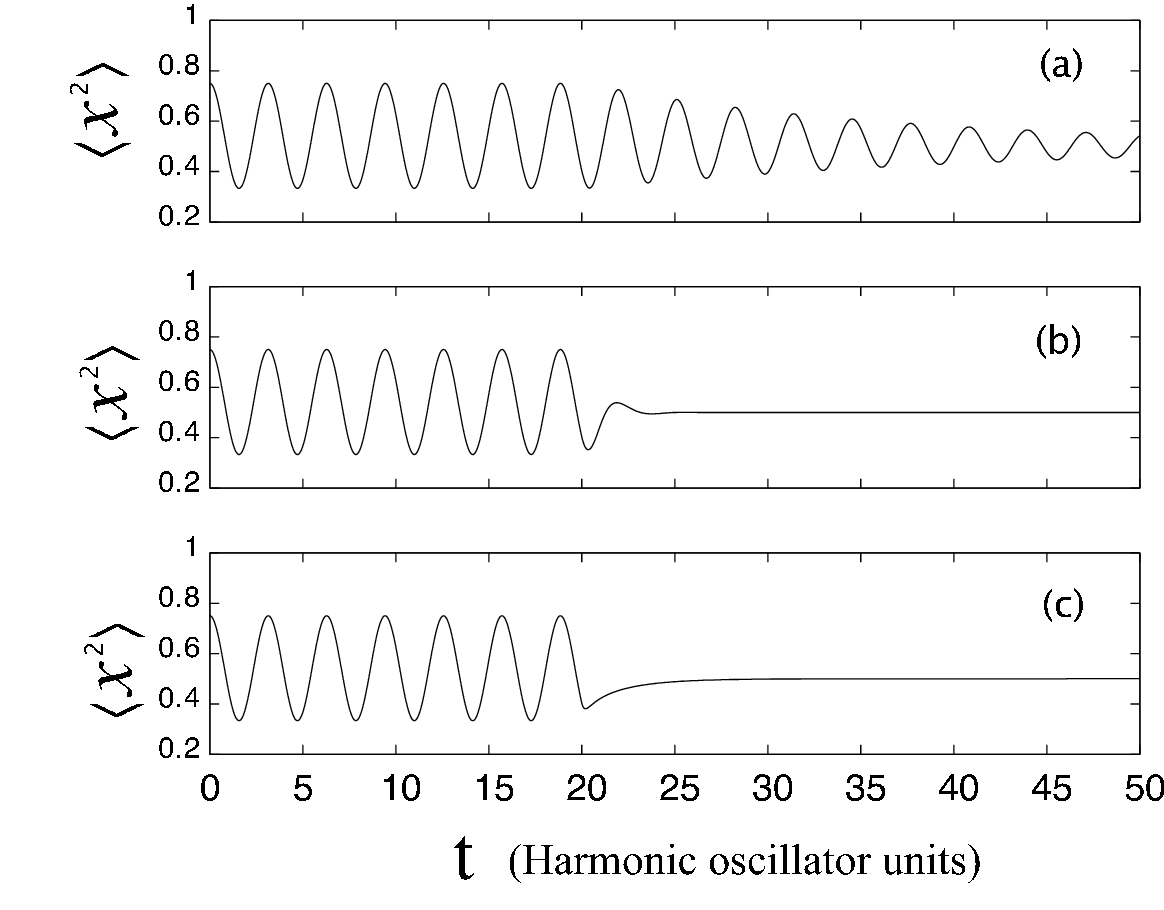}
\caption{\label{fig:crit_c2} Oscillations in condensate width versus time for (a) $c_2 = 0.05$; (b) $c_2 = 1$; (c) $c_2 = 5$.
It can be seen that (a) is underdamped, (b) is close to critical damping, and (c) is overdamped. $\langle x^2 \rangle$ and $t$ are measured in harmonic oscillator units.}
\end{figure}

Equation \ref{HOeom} was integrated numerically using a pseudospectral method with a fourth-order Runge-Kutta time step \cite{rk4ip} using MATLAB. The feedback initially turned off, and then switched on at time $t = 20$. Figure \ref{fig:crit_c2} shows how the oscillations in $\langle x^2 \rangle$ are damped for different values of $c_2$. It appears that critical damping occurs when $c_2 \approx 1$, and this value will be used for all subsequent calculations. 

\begin{figure}
\includegraphics[width=\columnwidth]{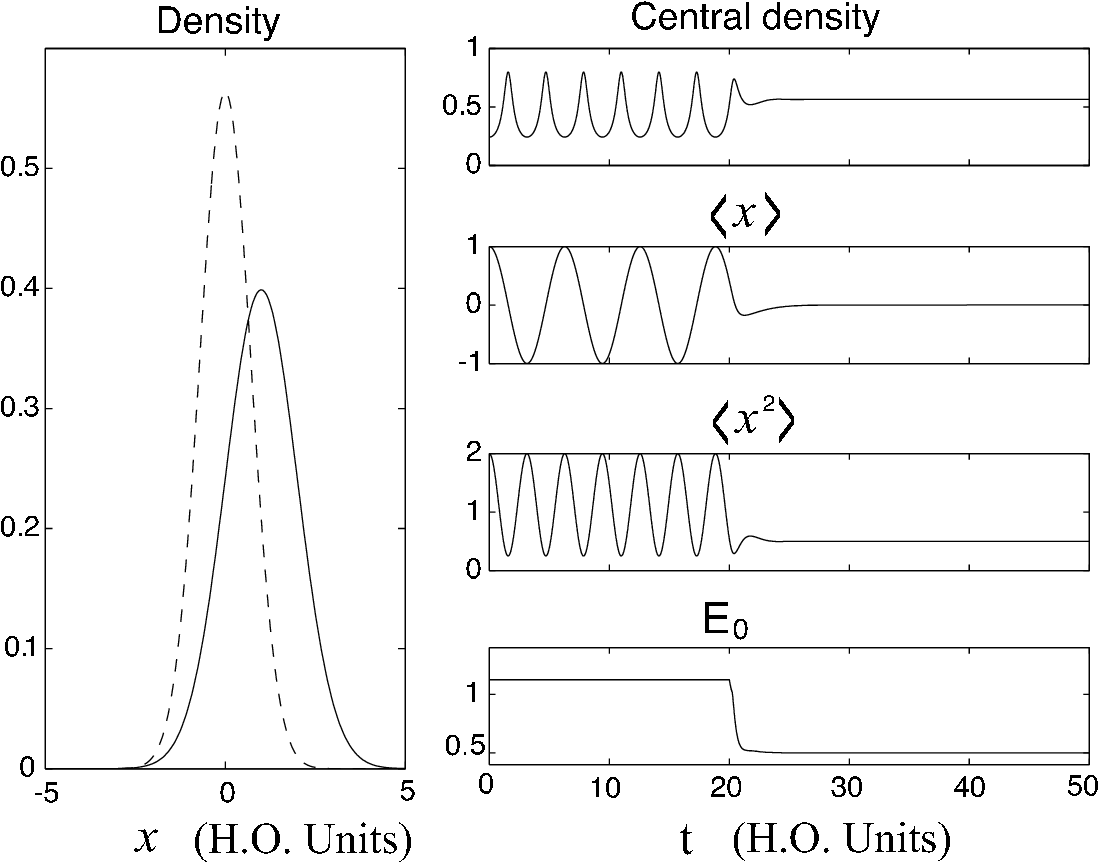}
\caption{\label{fig:linfeedback1} All quantities measured in harmonic oscillator units. Both modes of feedback working simultaneously on a system. The density profile of the initial condition is shown on the right with the solid black line, in comparison to the ground state density profile, indicated by the dashed line. The central density is the density at the point $x=0$. The energy is reduced to $E_0 = 0.5$, which is the ground state energy of the harmonic oscillator.}
\end{figure}

We next demonstrate that the two moments of feedback can be used together to reduce energy from the system. Figure (\ref{fig:linfeedback1}) shows the system initially in a non-stationary state. The feedback is turned on at time $t = 20$, and oscillations in $\langle x \rangle$ and $\langle x^2\rangle$ are quickly reduced. $E_0$ is reduced until it is $\frac{1}{2}$, which is the energy of the ground state wave function in a harmonic potential. 

In this particular example, the energy is reduced until the system is in the ground state.  Equation \ref{E0dot} shows that the energy will only be reduced when there are oscillations in $\langle x\rangle$ and $\langle x^2 \rangle$, so once the system is in a state where $\dot{\langle x \rangle}=0$ and $\dot{\langle x^2 \rangle}=0$, the feedback will no longer reduce the energy. Obviously, energy eigenstates will display no error signal, but these are not a problem as they are single mode and all expectation values of observables display no time dependence. Using the harmonic oscillator ladder operators ($\hat{a} = \frac{-i}{2}\bigl(\frac{\partial}{\partial x} +x\bigr)$, $\hat{a}^{\dag} = \frac{-i}{2}\bigl(\frac{\partial}{\partial x} -x\bigr)$) we can write $x= \frac{i}{\sqrt{2}}(\hat{a}-\hat{a}^{\dag})$. In the absence of error signals ($a_1(t)=a_2(t)=0$), we can use the Heisenberg equation of motion to calculate $\dot{\langle x\rangle}$ and $\dot{\langle x^2\rangle}$. By setting these equal to zero, we get a condition for our zero error signal states
\begin{equation}
\sum_{n=0}^{\infty} \sqrt{n+1}(\alpha^{*}_{n+1}\alpha_{n}e^{-it} +\alpha_n^{*}\alpha_{n+1}e^{it}) =0 \label{xdotnbasis}
\end{equation}
\begin{equation}
\sum_{n=0}^{\infty} \sqrt{n+1}\sqrt{n+2}(\alpha^{*}_{n+2}\alpha_{n}e^{-2it} -\alpha_n^{*}\alpha_{n+2}e^{2it}) =0 \label{xsqdotnbasis}
\end{equation}
where $|n\rangle$ are the energy eigenstates ($\hat{H}_0|n\rangle = E_n|n\rangle$), and $\alpha_n$ are their coefficients $|\psi\rangle = \sum_{n=0}^{\infty} \alpha_n e^{-i(n+\frac{1}{2})t} |n\rangle$. This shows us that there are an infinite number of non-stationary states that display no error signal. 

\begin{figure}
\includegraphics[width=\columnwidth]{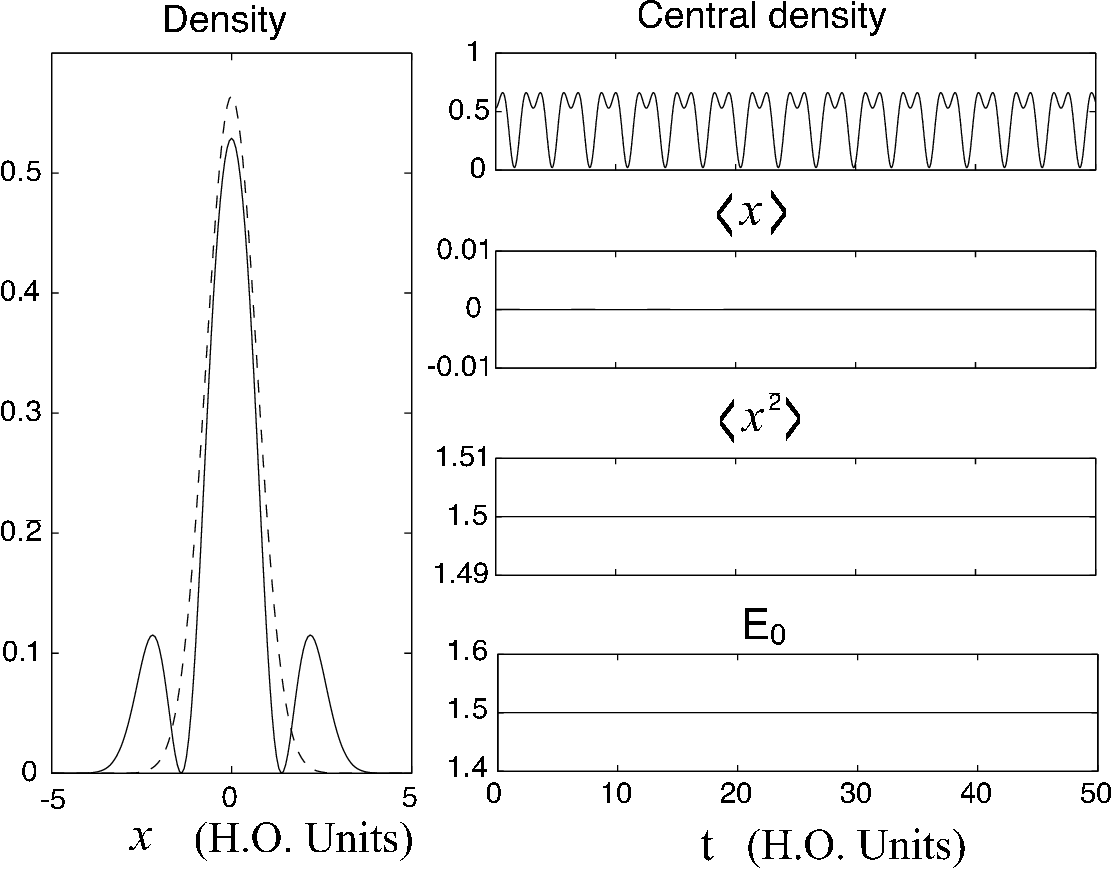}
\caption{\label{fig:linfeedback2} A state with no oscillations in $\langle x \rangle$ and $\langle x^2\rangle$. The feedback does nothing to reduce the energy as there is no error signal. }
\end{figure}

This result demonstrates that feedback using these controls will not always be effective, as the system may be attracted to one of these states rather than an eigenstate.  In these non-stationary states with no error signal, the energy will not be further reduced, and semiclassical fluctuations will continue. Figure(\ref{fig:linfeedback2}) shows an example of such a state.  It displays no oscillations in $\langle x\rangle$ and $\langle x^2 \rangle$, and the feedback does nothing to reduce the energy. The oscillations in the density at the centre of the trap are included to demonstrate that the condensate is dynamic. 
Obviously, our two error signals are insufficient to reduce dynamics fluctuations for the system in general. Our choice of error signal is governed by the controls we have available to us. We chose the curvature and position of the minimum of the harmonic potential as our controls as they are easy to manipulate in current experimental traps. In the next section we introduce a time dependent nonlinear interaction in an attempt to produce a feedback scheme that will remove all the semiclassical fluctuations. 

\subsection{Harmonic oscillator with a nonlinear control}
\begin{figure}
\includegraphics[width=\columnwidth]{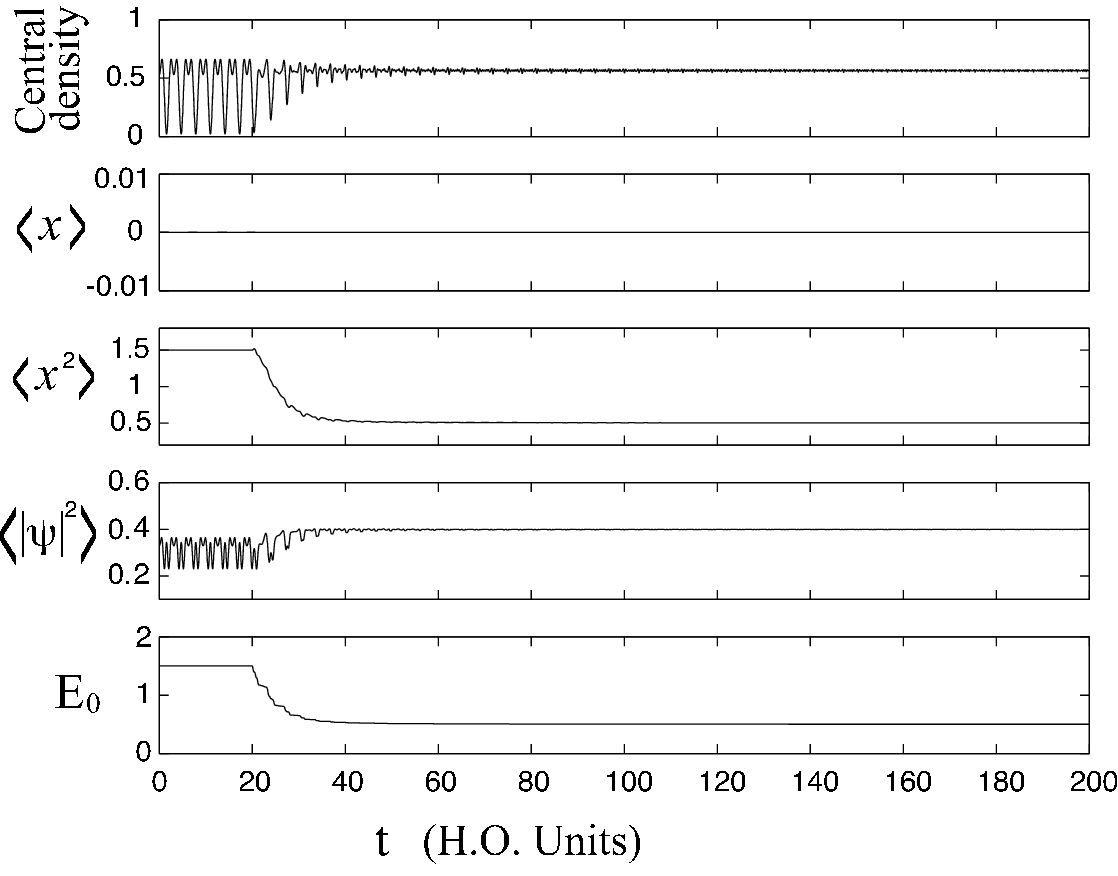}
\caption{\label{fig:linfeedback3} A condensate in the same initial state as \ref{fig:linfeedback2}, but feeding back using a time dependent nonlinear interaction with $u_1 = 5$ as well as the two trap parameters. In this case the additional error signal allows the feedback to reduce the energy until it is the ground state energy. The condensate number was normalized to unity for this example.}
\end{figure}

It is possible to tune the non-linear interaction between atoms in a Bose-Einstein condensate by controlling the magnetic field close to a Feshbach resonance\cite{Feshbach}.  In experimental systems, this is equivalent to controlling the bias magnetic field in a magnetic trap, or applying a constant magnetic field in an optical trap, and it has been achieved with considerable finesse in many recent experiments\cite{egBosenova}.  Adding a time dependent interaction between the atoms gives the equation of motion:
\begin{equation}
i\dot{\psi} = (\hat{T} + V_0 + a_1(t)x +a_2(t)x^2 +b_1(t)|\psi|^2)\psi \label{nHOeom}
\end{equation}
Setting $b_1(t) = u_1\dot{\langle|\psi|^2\rangle}$ in accordance with our feedback scheme will always reduce $E_0$. 
 Figure \ref{fig:linfeedback3} shows a system in the same initial state as figure \ref{fig:linfeedback2} but with the additional control. The additional error signal allows us to perturb the system from the stable state, and the energy is reduced to the ground state energy. 
We have demonstrated how we can use feedback effectively to remove energy from nonstationary states in the linear regime ($U_0=0$). In the follow section we look at the more physically realistic example of a Bose-Einstein condensate with a strong nonlinear interaction.

\section{Controlling a Bose-Einstein condensate with feedback}\label{section4}
\begin{figure}
\includegraphics[width=\columnwidth]{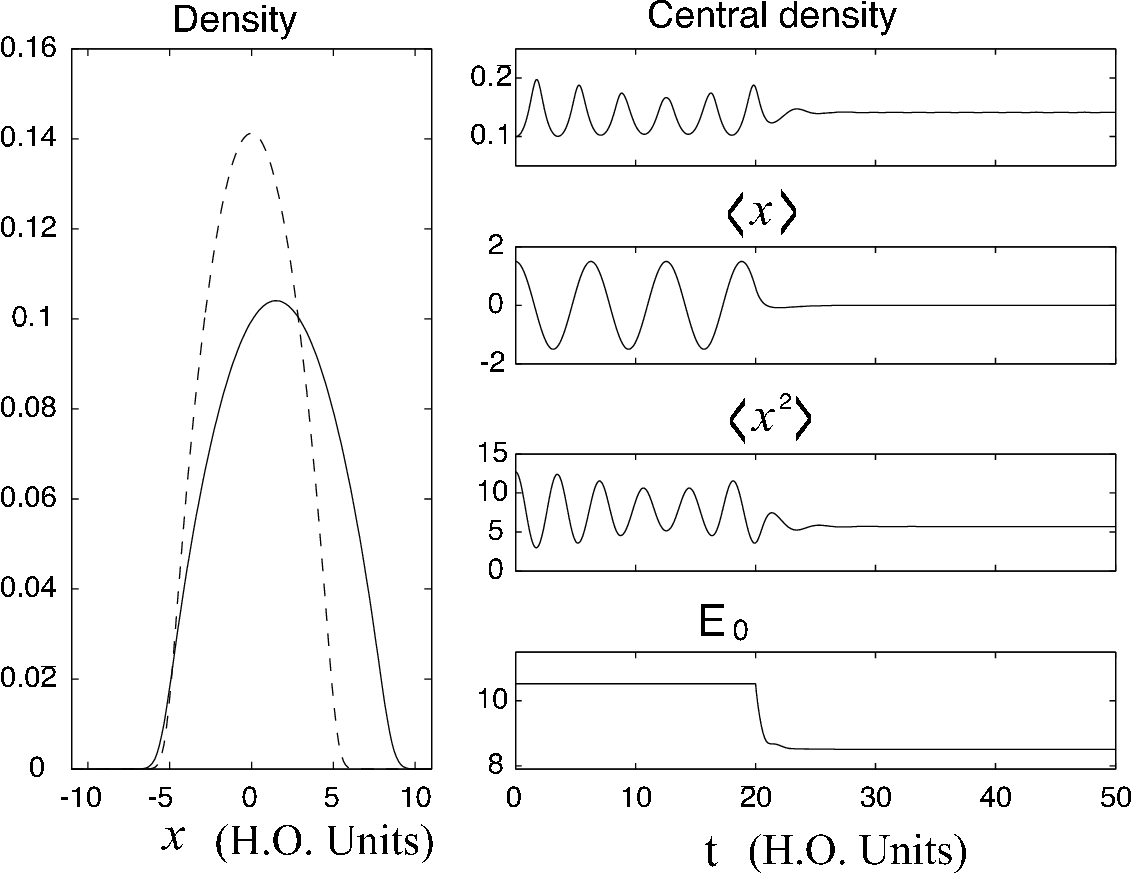}
\caption{\label{fig:gplinf1} Feedback on a condensate with a large nonlinear interaction ($U0 = 100$, condensate number normalized to unity) using $x$ and $x^2$ as our controls for the time dependent potential. The density profile of the initial state is shown on the left with a solid line, compared to the ground state with a dashed line. Oscillations in $\langle x \rangle$ and $\langle x^2\rangle$ are reduced and the energy is reduced to $E \approx 8.51$, which is the ground state energy. $c_2$ was chosen to be $0.05$ for this example.}
\end{figure}

\begin{figure}
\includegraphics[width=\columnwidth]{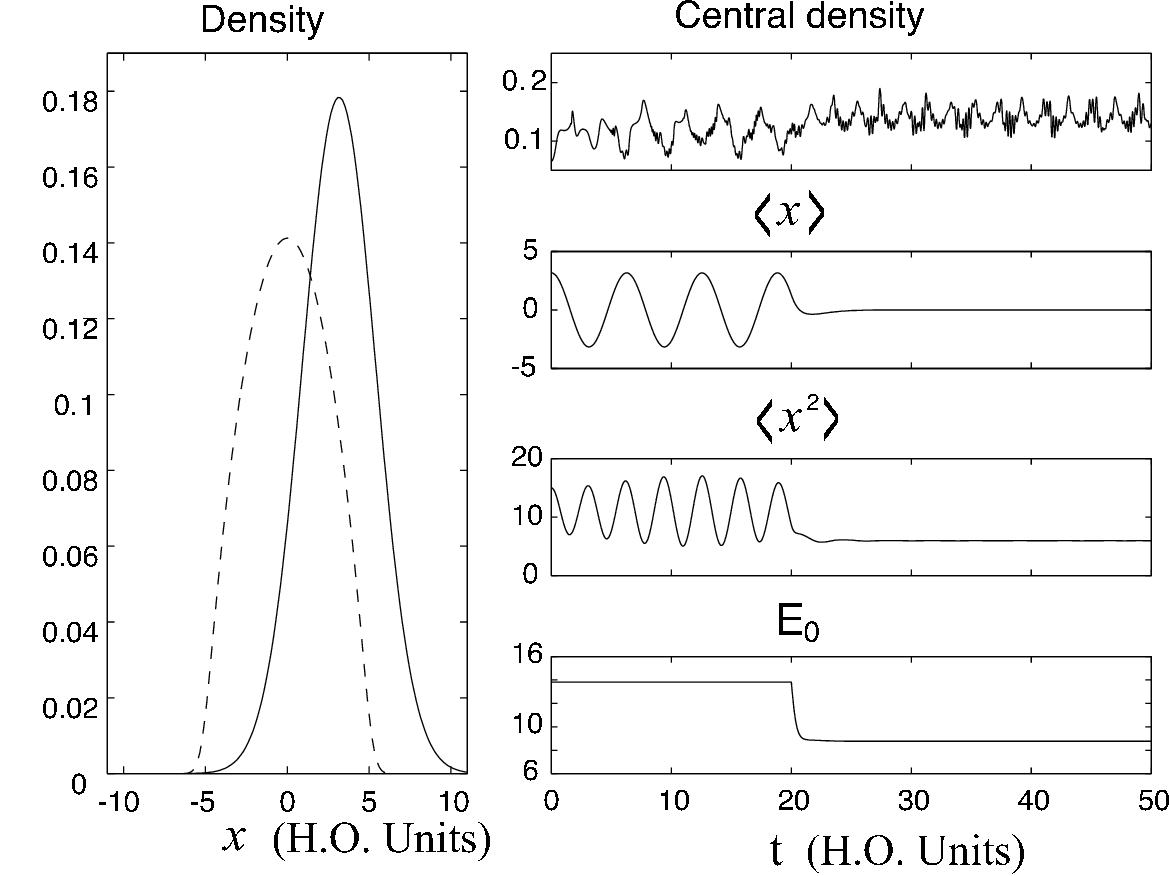}
\caption{\label{fig:gplinf2} Feedback on a condensate with a large nonlinear interaction ($U_0 = 100$, condensate number normalized to unity) in a different initial state. The feedback quickly removes energy from the two controlled modes, but energy in higher order excitations is more slowly reduced as it is coupled into the controlled modes via the nonlinear interaction.}
\end{figure}

\begin{figure}
\includegraphics[width=\columnwidth]{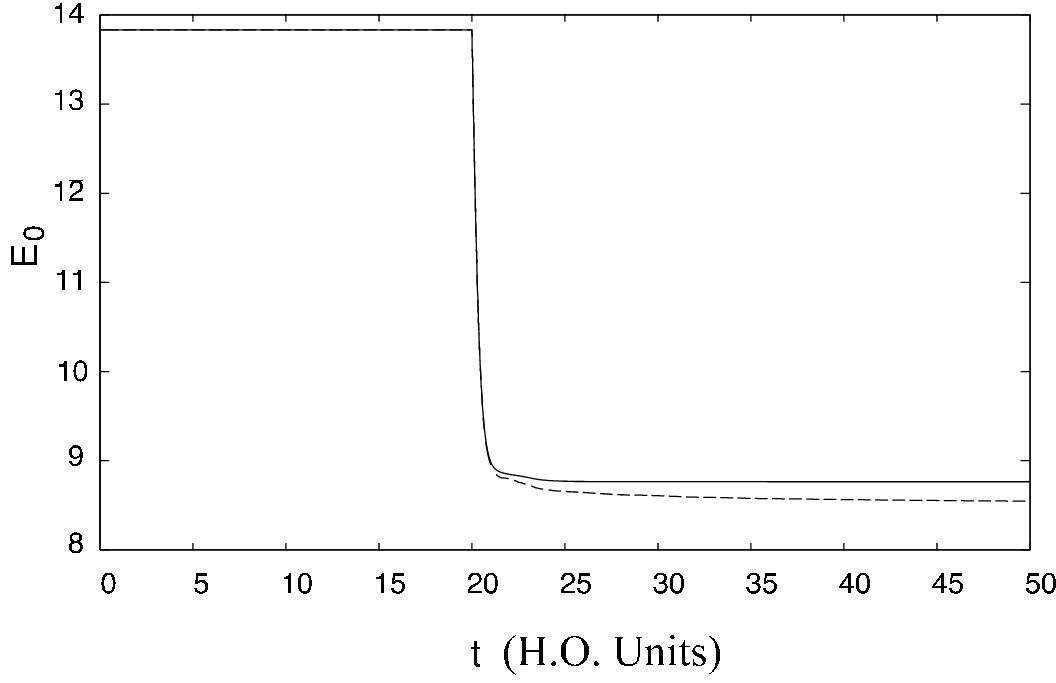}
\caption{\label{fig:gpenrgycompare} Comparison of energy reduction by feedback with and without the time dependent nonlinear interaction strength. The Solid line is $E_0$ for \ref{fig:gplinf2}, and the dashed line is $E_0$ with the time dependent nonlinear interaction included for $u_1=1000$.}
\end{figure}

We use as our next example the more realistic system of a Bose-Einstein condensate with strong interatomic interactions in a harmonic trap. We begin by just using the two trap controls as described in section \ref{section3} to reduce the energy. Figure \ref{fig:gplinf1} shows a condensate that is initially in an excited state, and the two modes of feedback reduce the energy until it is in the ground state. This is a special case, however, and figure \ref{fig:gplinf2} shows the feedback acting on a more general initial state.  The energy is quickly removed from the two controlled moments, but there is still energy left in higher energy excitations.  In contrast to the linear system, the motion in these higher moments is coupled into the controlled modes via the nonlinear interaction, and hence slowly reduced.  This is an inefficient process that may be allieviated by including the time dependent interaction strength as a third control.  Figure \ref{fig:gpenrgycompare} compares the results of using all three feedback controls on a BEC with non-zero interaction with the effects of using only the linear controls.  The use of the non-linear feedback dramatically accelerates the energy removal process after the rapid initial control due to the linear controls.

\section{Conclusion}
We have described a feedback scheme for reducing energy from a BEC in an arbitrary potential with an arbitrary set of controls.  This reduces the semiclassical fluctuations in the condensate, a process that will be essential for producing high quality atom lasers.  In the case of a linear harmonic oscillator with a modulated trapping potential, we demonstrated that energy can only be extracted from the moments in the motion corresponding to the moments present in the available controls.  The ability to modulate the nonlinear interaction between the atoms provides a feedback scheme that can control a far greater range of inital states.  Formally, any eigenstate will be unaffected by the feedback scheme, but as our scheme can only remove energy from the system, a slight perturbation will usually result in the system coming to steady state in a lower energy eigenstate.

In the case of a Bose-Einstein condensate with a large nonlinear interaction, there is already coupling between different modes of oscillations.  This means that each mode of feedback can remove energy from more than one mode of oscillation.  This indirect method of extracting energy from the higher modes is quite inefficient. Adding a nonlinear control improves the efficiency of the feedback because it directly removes energy from a larger range of modes.

\end{document}